\documentstyle[aps,preprint]{revtex}
\draft

\author{ Zafar Ahmed \\
Nuclear Physics Division, Bhabha Atomic Research Centre \\
Trombay, Bombay 400 085, India \\ zahmed@apsara.barc.ernet.in }
\title
{C, PT, CPT invariance of pseudo-Hermitian Hamiltonians}
\date{\today}
\begin{document}
\maketitle
\begin{abstract}
We propose construction of a unique and definite metric ($\eta_+$),
time-reversal operator (T) and an inner product such that
the pseudo-Hermitian matrix Hamiltonians are C, PT, CPT invariant and
PT(CPT)-norm is indefinite (definite). Here, P and C denote the generalized
symmetries : parity and charge-conjugation respectively. The limitations
of the other current approaches have been brought out.
\end{abstract}
\vspace {.2 in}
\section {Introduction : PT-Symmetry and Pseudo-Hermiticity}
\par Last few years have witnessed a remarkable development wherein
the discrete symmetries of a Hamiltonian seem to decide if the  eigenspectrum
will be real. It has been conjectured [1] that Hamiltonians possessing symmetry
under the combined transformation of parity (P: $x\rightarrow -x$) and time-reversal
(T : $ i\rightarrow -i$) will have real discrete spectrum provided the eigenstates
are also simultaneous
eigenstates of PT. Interesting situations are those where P and T are individually
broken. An overwhelming number of evidences supporting the conjecture are available.
[1-7].
\par The real eigenvalues of a PT symmetric Hamiltonian are found connected
with a more general property of the Hamiltonian namely the pseudo-Hermiticity.
The concept of pseudo-Hermiticity was developed in 50s-60s [9] following
definition of a distorted definition of inner product $\langle \eta \rangle$ [8], $\eta$ is
called a metric. A Hamiltonian is called pseudo-Hermitian, if it is such that
\begin{equation}
\eta H\eta^{-1}=H^\dagger.
\end{equation}
The eigenstates corresponding to real eigenvalues are $\eta$-orthogonal
and eigenstates corresponding to complex eigenvalues have zero $\eta$-norm (2).
Identifying $\eta$ for a non-Hermitian Hamiltonian when it has real eigenvalues
is very crucial. Most of the PT-symmetric Hamiltonians having real eigenvalues
have recently been claimed to be P-pseudo-Hermitian, and several other interesting
results have been derived [10].
Several non-Hermitian Hamiltonians of both types PT-symmetric and non-PT-symmetric
possessing real spectrum have been identified as pseudo-Hermitian under
$\eta=e^{-\theta p}$ and $e^{-\phi(x)}$ [11]. Some more interesting developments
relate to weak-pseudo-Hermiticity [12], pseudo-anti-Hermiticity [13] and construction
non-PT-symmetric (pseudo-Hermitian) complex potential potentials  having real
eigenvalues [14] A new pseudo-unitary group and Gaussian-random pseudo-unitary
ensembles of pseudo-Hermitian matrices have been proposed [15]. This development
gives rise to new energy-level distributions which are expected to represent
the spectral fluctuations of PT-symmetric systems.
\par Most interesting feature of the eigenstates of such Hamiltonians is the
indefiniteness [5-8] (positivity-negativity) of the norm which is the consequence
of the $\eta$-inner product [8]
\begin{equation}
\langle \Psi_m |\eta \Psi_n \rangle = \epsilon_n \delta_{m,n},
\end{equation}
where $\epsilon_n(=\pm 1)$ is indefinite (positive-negative). Recall, that
the usual norm in Hermiticity is $\langle \Psi_n | \Psi_n \rangle$ positive-definite.
Currently, the negativity of the PT-norm has been proposed to indicate the
presence of a hidden symmetry called C which mimics charge-conjugation symmetry
(${\cal C}$) [17].
It has been claimed that CPT-norm will be positive definite.
An interesting scope for PT-symmetric quantum field theory has been argued.
The construction
of the new involutary operator C has been discussed. A $2\times 2$ matrix
Hamiltonian which is actually pseudo-Hermitian with real eigenvalues has been employed
and by constructing P=$\eta$, T=$K_0$ and a CPT-norm the, the novel proposal has
been illustrated [16]. $K_0$ represents complex-conjugation operator; e.g.
$K_0 (A B)=A^\ast B^\ast$. Though sufficient and consistent for their assumed
model of non-Hermitian Hamiltonian, let us remark that these constructions
are too simple to work in general.
\section {Current developments and motivation}
\par The next related development [18] caters to the construction of generalized
involutary operators C,P,T from the bi-orthonormal [8,14] basis ($\Psi,\Phi)$ of the
pseudo-Hermitian Hamiltonian with real eigenvalues. In doing so, the well
developed machinery of pseudo-Hermiticity has been aptly utilized.
This development, however, does not dwell upon the
negativity of the PT-norm and invoking of C for the positive-definiteness of
the CPT-norm. In this approach, the search of various symmetries of H and their
identification as C,PT or CPT has been proposed. Despite, obtaining a curiously
different definition of T other than the simple $K_0$ [16], this dichotomy has
neither been remarked nor resolved. Also, despite this incompatibility a similar
definition of the CPT-inner product [16] has been adopted [18].
\par Further, when a Hamiltonian is Hermitian and of the type ${\cal H}=p^2/(2m)
+V(x)$ by adopting the definition of C [16] which becomes P now due to Hermiticity,
it has been claimed that Hermitian operators, ${\cal H}$ have parity P and they are
PT-invariant. It may be noted that, the definition of P proposed in [19] is
identical to the definition of generalized parity proposed in [18] when the
Hamiltonian becomes Hermitian.
\par While following these developments one very strongly feels
that a Hermitian Hamiltonian ought to be P, T, PT, and CPT invariant. The PT (CPT)-norm
ought to be indefinite (definite). Also the eigenstates of H should display
the orthonormality consistent with the definition of norm under the same inner
product. These primary contentions do however not meet in either of the approaches
[16,18]
\par In fact, these expectations have been met lately, not without incorporating
a generalized definition of T [20] {\it a la}, discarding T$=K_0$ [16] and proposing
an inner product [20]. In this Letter, we propose further extension of these [20]
definitions so as to bring consistency in proposing the C, PT, and CPT invariance of
a pseudo-Hermitian Hamiltonian (real eigenvalues) definiteness of CPT-norm and
the indefiniteness of PT-norm. In our studies, we prefer the use of matrix
notations and matrix models of Hamiltonians. Recall that in case of Hermiticity, for
the usual stationary states the three modifications $\Psi(x), \Psi^\ast(x)$ and
$\Psi^\dagger(x)$ usually coincide. However, in matrix notations, we have four
distinct modifications of the state these are $\Psi,\Psi^\ast$  (complex-conjugate)
,$\Psi^\prime$ (transpose), $\Psi^\dagger$ (transpose and complex-conjugate).
This makes the matrix notations more general, unambiguous and unmistakable.
\section {Pseudo-Hermitian matrices : a unique and definite metric}
\par Let us notice the non-Hermitian complex matrix, $H$, given below
admitting real eigenvalues $E_{0,1}=a \pm \sqrt{bc},$ when $b c > 0$.
We find that there exist four metrics $\eta_i$ under which H is pseudo-Hermitian
\begin{eqnarray} H =\left [\begin {array} {cc} a & -ib \\ ic & a \end {array} \right],
~~\eta_1 =\left [\begin {array} {cc} 0   & -i\\ i & 0
\end {array} \right],~~\eta_2 =\left [\begin {array} {cc} r^2 & -s \\ s & 1
\end {array} \right],~~\eta_3 =\left [\begin {array} {cc} r   & 0 \\ 0 & 1/r
\end {array} \right],
\eta_4 =\left [\begin {array} {cc} 0   & -1 \\ 1 & 0 \end {array} \right]
\end{eqnarray}
Here $r=\sqrt{c/b}$ and $s$ is in general an arbitrary complex number,
indicating that a metric need not necessarily be Hermitian. These $\eta_1$
(Pauli's $\sigma_x$) and $\eta_{2,3,4}$ have ,in fact, been found by crude
algebraic manipulations demonstrating that metric $\eta$ is non-unique as
informed earlier [10]. Furthermore,
if $\eta_1$ and $\eta_2$ are found then infinitely many metrics can be constructed
as $\eta=(c_1\eta_1+c_2\eta_2)$ provided $\eta$ is invertible.
On one hand, the four metrics
given above (3) do provide several operators $F_{i,j}=\eta_i \eta^{-1}_j, i \ne j=1,4$
which by commuting with $H$ bring out its hidden symmetries [10].
In fact, the currently discussed C, PT, and CPT symmetries shall be seen connected
to $F_{i,j}$ in the examples to follow in the sequel.
On the other hand, the non-uniqueness of $\eta$ apart from its indefiniteness
may be undesirable as the metric determines the expectation values of various
operators as $\langle \Psi| A\eta \rangle$.
We state and prove the following theorem which helps us in fixing a unique
and definite metric. This could be seen as a method to find at least one metric
under which a given matrix is pseudo-Hermitian. \\
{\bf Theorem  :}\\
If a diagonalizable complex matrix $H$ admits real eigenvalues $(E_1,
E_2,...E_n)$ and $D$ is its diagonalizing matrix then $H$ is
$\eta$-pseudo-Hermitian, where $\eta=(DD^\dagger)^{-1}$.
Converse of this also holds.\\
{\bf Proof :} Let
\begin{mathletters}
\begin{equation}
D^{-1} H D =Diag [E_1,E_2,...,E_n]
\end{equation}
\begin{equation}
\Rightarrow D^{-1} \eta^{-1} \eta H \eta^{-1} \eta D = Diag [E_1,E_2,...,E_n]
\end{equation}
Invoking the pseudo-Hermiticity (1), we write
\begin{equation}
D^{-1} \eta^{-1} H^\dagger \eta D = Diag [E_1,E_2,...,E_n]
\end{equation}
The transpose-conjugation of Eq. 3(a) yields
\begin{equation}
D^\dagger H^\dagger (D^{-1})^\dagger = Diag [E_1,E_2,...,E_n]
\end{equation}
\end{mathletters}
Upon comparing last two equations, we get $D^{-1} \eta^{-1}=D^\dagger$
and $\eta D= (D^{-1})^\dagger$ which imply  $\eta=(DD^\dagger)^{-1}$.
$ \Box $ \\
When H is Hermitian, D will be unitary and we get $\eta=I$ as a special case.
In general, D will be pseudo-unitary : $D^\dagger=\delta D^{-1}
\delta^{-1}$ [8,15]. w.r.t. some metric $\delta$ which may not be same as
$\eta$.\\
{\bf Proof (Converse) :} Let
\begin{mathletters}
\begin{equation}
D^{-1} H D= Diag[E_1,E_2,E_3.....,E_n].
\end{equation}
\begin{equation}
\mbox {and} ~~~(DD^{\dagger})^{-1} H (DD^\dagger) = H^\dagger
\end{equation}
\begin{equation}
\Rightarrow (D^{\dagger})^{-1} (D^{-1} H D ) D^\dagger = H^\dagger.
\end{equation}
\begin{equation}
\Rightarrow (D^{\dagger})^{-1} (Diag[E_1,E_2,E_3.....,E_n] )
D^\dagger = H^\dagger.
\end{equation}
By taking transpose-conjugate on both the sides, we have
\begin{equation}
\Rightarrow D (Diag[E_1,E_2,E_3.....,E_n] )^\dagger
D^{-1} = H.
\end{equation}
By left (right) multiplying by D (D$^{-1}$) on both the sides, we get
\begin{equation}
\Rightarrow (Diag[E^\ast_1,E^\ast_2,E^\ast_3.....,E^\ast_n] )= D^{-1} H D.
\end{equation}
\end{mathletters}
Eq. (5e) and (5f) imply nothing but the reality of eigenvalues.   $\Box$\\
Similarly, when all the eigenvalues are complex conjugate and $D$ is
the diagonalizing arranged such that complex conjugate pairs remain together
then it can be proved that $\bar{\eta}=(D S D^\dagger)^{-1}$, where $S$ is Pauli's
$\sigma_x$, when $H$ is $2 \times 2$ otherwise when $H$ is $2n \times 2n,$
$S$ is block-diagonal matrix : $S=Diag [\sigma_x,\sigma_x, \sigma_x,....\sigma_x]$.
We now denote and state thus obtained metric as
\begin{equation}
\eta_+=(D D^\dagger)^{-1},
\end{equation}
to actually see that the indefinite norm (2)
\begin{equation}
N_{\eta+}=\Psi^\dagger \eta_+ \Psi=\Psi^\dagger (DD^\dagger)^{-1}\Psi=
\Psi^\dagger {D^\dagger}^{-1} D^{-1}\Psi=(D^{-1}\Psi)^\dagger (D^{-1} \Psi)=
\chi^\dagger \chi >0.
\end{equation}
is now positive definite.
Finding eigenvalues, eigenvectors and diagonalizing matrix is a standard
exercise. In that the theorem
stated and proved above is indeed an attractive proposal to find the metric
for a given complex non-Hermitian matrix admitting real eigenvalues under which
it is pseudo-Hermitian. However, by multiplying the columns (rows) by arbitrary
constants we can get many diagonalizing matrices say $D_j$ and this
would give rise to
as many metrics say $\eta_j$ under which $H$ will be pseudo-Hermitian.
For the sake of uniqueness, one may only use $\eta$-normalized (2) eigen-vectors
to construct $D$. Earlier, it has been proved that if a pseudo-Hermitian
Hamiltonian, H, has real eigenvalues then there exists and operator O such that
H is pseudo Hermitian under: $(OO^\dagger)$
[10] and $(OO^\dagger)^{-1}$ [12]. Another, form for $\eta_+$ in terms of the eigenvectors
has also been proposed [18].
\section {Construction of C,P,T and proposal of an inner product}
When pseudo-Hermitian Hamiltonian (1) has real eigenvalues, we have [8]
\begin{equation}
H\Psi_N=E_n\Psi,~~~H^\dagger \Phi_n=E_n \Phi_n,~~~ \Phi=\eta \Psi,
\end{equation}
$(\Psi_n,\Phi_n)$ are called bi-orthonormal
basis and $\Phi=\eta \Psi$. We have also witnessed in the example above (3) that
several metrics could be obtained under which a given H is pseudo-Hermitian.
Let us stress that this interesting practical experience remains elusive in
several formal definitions.
Let us examine the properties of the metrics obtained in (3). The metric
$\eta_1$ is involutary ($U^2=1$). The metrics $\eta_1,\eta_3,\eta_4$ are
Hermitian, unitary and simple ($det U=1$). The metrics $\eta_3,\eta_4$ are
real-symmetric. The metric $\eta_2$ very importantly is non-Hermitian in
general. The metrics $\eta_1, \eta_4$ are (constant) disentangled with the
elements of H and we call them as ${\it secular}$ [15].
It will be very interesting to investigate whether or not one can
always find an involutary and ${\it secular}$ metric for an arbitrary
pseudo-Hermitian matrix.
The interesting exposition [10] that most of the known PT-symmetric
Hamiltonians are actually P-pseudo-Hermitian is very valuable in order to
connect pseudo-Hermiticity with P and T and hence to possible physical
situations [15]. Once, the involutary metric is found it will be fixed
for the definition of orthonormality (2) and we will assume it to represent
the generalized P. This {\it ad-hoc} strategy also seems to have been adopted
in [16]. Therefore, the question of a definition to construct P again,
from the bi-orthonormal basis $(\Psi,\Phi)$ either does not arise or will
yield P=$\eta$, eventually.
\par Here, one very important remark is in order : in the recent
works on pseudo-Hermiticity, the indefiniteness of the $\eta-$norm
(or orthonormality) has not been realized and this has given rise to an
assumption that {\it somehow} $\Phi^\dagger_m \Psi_n$ is positive-definite
(e.g., Eqs. (11,12) in [10], Eqs. (5,6) in [12], Eq. (7) in [13]).
Consequently, representations of $I$ (the completeness) in terms of $(\Psi,\Phi)$,
for instance, for two level matrix Hamiltonian, has been given as
$(\Psi_0 \Phi^\dagger_0 + \Psi_1 \Phi^\dagger_1).$ Though, known earlier [3-9]
, however, the indefiniteness of the norm is centrally consequent to the
novel identification of charge-conjugation symmetry by Bender et. al.[16].
\par Thus having fixed $\eta$ for H, we find $\eta$-normalized (2) eigenvectors
$\Psi_n$. These, normalized eigenvectors are used to construct the diagonalizing
matrix D and $\eta_+$ (6) which are unique only under the fixed $\eta$.
We obtain another basis $\{\Upsilon_n\}$ as
\begin{equation}
\Upsilon_n=\eta_+ \Psi_n,
\end{equation}
which by construction (see (7)) is such that
\begin{equation}
\Psi^\dagger_m \Upsilon_n =\delta_{m,n}.
\end{equation}
In the spirit of [18], we propose to construct P as
\begin{equation}
P=\sum_{n=0}^{N} (-1)^n \Psi_n \Psi^\dagger_n,
\end{equation}
such that P$\Upsilon_n=(-)^n\Psi_n$, implying that neither of $\Psi_n,\Upsilon_n$
are the eigenstates of parity as it should be.
We define the anti-linear time-reversal operator T as
\begin{equation}
T=\left (\sum_{n=0}^{N} \Upsilon_n \Upsilon^\prime_n \right) K_0
\end{equation}
such that T$\Psi_n=\Upsilon_n$ and we further have
\begin{equation}
PT=\left (\sum_{n=0}^{N} (-)^n \Psi_n \Upsilon^\prime_n \right) K_0,
\end{equation}
such that PT$\Psi_n=(-)^n\Psi_n$.
We adopt the definition of C as proposed in [18]
\begin{equation}
C=\sum_{n=0}^{N} (-1)^n \Psi_n \Upsilon^\dagger_n,~~~ \mbox{where}
~~~\sum_{n=0}^{N} \Psi_n \Upsilon^\dagger_n =1
\end{equation}
such that C$\Psi_n=(-)^n\Psi$.
Next using (13) and (14) the symmetry operator CPT takes the form
\begin{equation}
CPT=\left (\sum_{n=0}^{N} \Psi_n \Upsilon^\prime_n \right) K_0,
\end{equation}
such that CPT $\Psi_n=\Psi_n.$
The following involutions
\begin{equation}
(CPT)^2=(PT)^2=C^2=1
\end{equation}
always hold. However, we get
\begin{equation}
T^2=P^2, ~~~\mbox{iff}~~~ (-)^{m+n} \Psi^\dagger_m \Psi_n =
\Upsilon^\dagger_m \Upsilon_n.
\end{equation}
When the Hamiltonian is Hermitian, P and T have been proved to be involutary [20].
However, for pseudo-Hermitian Hamiltonian this becomes conditional.
In Eq. (87) of [18], the above condition is suggested to be ensuring that
P and T are involutary. Let us remark that this condition only ensures that
$P^2=T^2$. Further, since we choose P to be involutary and so will T be.
We find that the following commutation relations
\begin{equation}
[H,C]=[H,PT]=[H,CPT]=0,~~~\mbox{and~~~} [H,P] \ne 0 \ne [H,P]
\end{equation}
displaying the invariance and non-invariance of the Hamiltonian.
We now define a X-inner product as
\begin{equation}
(X \Psi_m)^\dagger \Upsilon_n = (X \Psi_m)^\dagger \eta_+ \Psi_n=\epsilon_n
\delta_{m,n},
\end{equation}
where $\epsilon_n(=\pm 1)$ is indefinite. Consequently, the X-norm as
\begin{equation}
N_{X,n}=(X \Psi_n)^\dagger \Upsilon_n = (X \Psi_n)^\dagger \eta_+ \Psi_n.
\end{equation}
Here $X$ represents the symmetry operators such as C, PT, and CPT constructed above,
such that $[H,X]=0.$
Since $X \Psi_n=\epsilon_n \Psi$, $\epsilon_n$ is real, the X-inner product
in view of (7) will be real-definite.
\section {examination of the other current approaches}
Let us examine the inner products defined in [16] and [18].
The inner product (Eqs.(5,12,22) in [16]) in our notations reads as
\begin{equation}
(X \Psi_m)^\prime \Psi_n,
\end{equation}
which is not real-definite in general, noting the fact that $\Psi_n$
are eigenvectors
over a complex field (the elements of these vectors are complex).
The same shortcoming of not being real-definite applies to the inner product
analysed and proposed in (Eq. (75) in [18]) which would read as
\begin{equation}
(X \Upsilon_m)^\prime\Psi_n
\end{equation}
We have earlier [20] proved and illustrated that the definition of the inner product
(21) [16] does not let the energy-eigenstates of the Hermitian H to be orthogonal.
We would like to claim
that our definition of the X-inner product proposed here is most general
and consistent so far [3-9,16,18-20], for the PT-symmetric or pseudo-Hermitian Hamiltonians.
\par Let us now appreciate how despite the inner product (21) not being real-definite
in general, the physically intriguing and also consistent claims of C, PT, and CPT
invariance of H and definiteness of CPT-norm could have been made.
The eigenvectors of a pseudo Hermitian matrix are naturally $\eta$-orthogonal (2)
let us remark that H in [16] is pseudo-Hermitian under $\eta=\sigma_x$, which has been
chosen to be P. In fact, H (Eq.(14) in [16]) is a special example, where the elementary
($\psi_n$) eigenvectors are also {\it incidentally} orthogonal as $\psi^\prime_0 \psi_1=0,$
in addition to the $\eta$-orthogonality : $\psi^\dagger_0 \eta \psi_1=0$.
Therefore, the concept and method of pseudo-Hermiticity which promises generality
could be relaxed here [16].
Next, these eigenvectors are to be multiplied by suitable factors to obtain
the relevant useful basis, $\{\Psi_n \}$, such that
$PT \psi_n=\eta K_0 \Psi_n=(-)^n \Psi_n$.
We would like to add one more such instance. where this method could succeed
again is the following
\begin{eqnarray} H =\left [\begin {array} {cc} a-c & ib \\ ib & a+c
\end {array} \right],
~~~\eta_ =\left [\begin {array} {cc} 1   & 0\\ 0 & -1 \end {array} \right]=P,
~~~\psi_0 =\left [\begin {array} {c} 1 \\ -ir  \end {array} \right],
~~~\psi_1 =\left [\begin {array} {c} 1 \\ -i/r  \end {array} \right],
\end{eqnarray}
where, we again have $\psi^\prime_0 \psi_1=0$, besides the $\eta$-orthogonality (2).
The eigenvalues are $E_{0,1}=a\pm\sqrt{c^2-b^2}, r={c+\sqrt{c^2-b^2} \over b}$
these are real as long as $c^2 >b^2$.
The illustrations ${\bf I_1, I_2}$ given below are also aimed at citing examples
where the approach taken in [16] does not work. It is , however, worth mentioning
that the prescriptions suggested in Section IV, which are in keeping with the spirit
of the approach in [18] {\it sans} the inner-product dwfined there and T, works
for both the examples :
one in [16] and the other discussed above in (23). The most notable failure of
the approach  in [16] has already been reported in [20] when it is applied back
to Hermiticity.

\section {illustrations}
The definitions for the construction of P,T,C, though general, certain features
can still not be proved.
For instance whether C and P will always not  commute. Whether P and T
will always commute. When a complex (non-Hermitian) matrix Hamiltonian
having real eigenvalues has P, which is not involutary will we get an
involutary T ? In this regard, simple doable examples are desirable.
In the following we present two illustrations to throw some more light
for the un-answered questions stated here.
\par  Without loss of generality, we take $2\times 2$ matrix Hamiltonians [15]
and construct P,T,C as per Eqs. (11), (12) and (14) as
\begin{equation}
P=\Psi_0 \Psi_0 - \Psi_1 \Psi_1,~~T=\left(\Upsilon_0 \Upsilon^\prime _0 +
\Upsilon_1 \Upsilon^\prime_1\right) K_0,~~C=\Psi_0\Upsilon_0-\Psi_1\Upsilon_1,
\end{equation}
for short. In illustration : ${\bf I_1}$, we take up the same Hamiltonian
as given in (3), here the fundamental metric (P) is involutary and in
illustration : ${\bf I_2}$, it is kept non-involutary.      \\
${\bf I_1:}$\\
We take pseudo-Hermitian Hamiltonian, H, and the fundamental metric,
$\eta(=\eta_1)$, from (3). The $\eta$-normalized eigenvectors are
\begin{eqnarray}
\Psi_0=\sqrt{2\over r}\left [ \begin{array}{c} -i/r \\ 1  \end{array}\right],
\Psi_1=\sqrt{2\over r}\left [ \begin{array}{c} 1/r \\ -i  \end{array}\right]
\end{eqnarray}
One can readily check that $\Psi^\dagger_0 \eta \Psi_1=0,$
but $\Psi^\prime_0 \Psi_1 =-i{1+r^2 \over 2r} \ne 0$ for the approach
[16] to work here.
Following section IV , we obtain P,T, and $\eta_+$  as
\begin{eqnarray} P =\left [\begin {array} {cc} 0 & -i \\ i & 0 \end {array} \right],
~~T =\left [\begin {array} {cc} 0   & -i\\ -i & 0 \end {array} \right] K_0,
~~\eta_+ =\left [\begin {array} {cc} r   & 0 \\ 0 & 1/r \end {array} \right],
\end{eqnarray}
Notice that P turns out to be the same as $\eta_1$-the chosen fundamental
metric. The symmetry operators C, PT and CPT are
\begin{eqnarray} C =\left [\begin {array} {cc} 0 & -i/r \\ ir & 0 \end {array} \right],
~~PT =\left [\begin {array} {cc} -1   & 0 \\ 0 & 1 \end {array} \right] K_0,
~~CPT =\left [\begin {array} {cc} 0 & -i/r \\ -ir & 0 \end {array} \right] K_0,
\end{eqnarray}
The symmetry operator C could be checked to be identical to $\eta_1 \eta^{-1}_3.$
(see (3)), demonstrating how two distinct metrics combine to yield a hidden
symmetry of the Hamiltonian. In addition to the general results stated above,
we get $(CP)^{-1}=PC=\eta_+$, in actual ${\cal CPT}$-invariance ${\cal C},
{\cal P}$ do commute [17]. We also confirm the commutation of P and T and the
involutions : $T^2=P^2=1.$ Similar, experience can be had by studying the model of
[16] and (23). Interestingly, the fundamental metrics in all these cases are the Pauli's
matrices which are involutary, Hermitian, unitary, simple and also {\it secular}.
\\
${\bf I_2 :}$\\
In the following, let us now take an example where the fundamental metric
is only Hermitian and {\it secular} as it does not affect the eigenvalues
: $E_{0,1}={1 \over 2}[(a+b)\pm \sqrt{(a-b)^2+4c^2}].$ We introduce $\theta=
{1 \over 2} \tan^{-1} {2c \over a-b}.$
\begin{eqnarray} H =\left [\begin {array} {cc} a & -ic/x \\
ic x & b \end {array} \right],
~~\eta =\left [\begin {array} {cc} x   & 0\\ 0 & 1/x \end {array} \right],
~~\Psi_0 =\sqrt{x}\left [\begin {array} {c} \cos\theta/x \\ i \sin \theta
\end {array} \right],
~~\Psi_1 ={1 \over \sqrt{x}}\left [\begin {array}{c} i \sin\theta \\ x \cos \theta
\end {array} \right],
\end{eqnarray}
Check that the states are only $\eta$-orthogonal and the condition
$\Psi^\prime_0 \Psi_1 = i\sin\theta(1+x^2)/x \ne 0$ like in ${\bf I_1}$
and unlike in Section V, is not met here.
We construct P, T, C as
\begin{eqnarray}
P =\left [\begin {array} {cc} {\cos 2\theta \over x} & -i \sin 2 \theta
\\i \sin 2 \theta & -x\cos 2\theta \end {array} \right],
~T =\left [\begin {array} {cc} x\cos 2\theta & i \sin 2 \theta
\\i \sin 2 \theta & {\cos 2\theta \over x} \end {array} \right]K_0,
~C =\left [\begin {array} {cc} \cos 2\theta & -{i \sin 2 \theta \over x}
\\i x \sin 2 \theta & -\cos 2\theta \end {array} \right]
\end{eqnarray}
and $\eta$ is returned as $\eta_+$.
Very interestingly, P is different from the fundamental metric $\eta$.
Since this fundamental metric is definite giving $\Psi^\dagger_n \eta \Psi_n=+1,$
the construction of $\eta_+$ as per (6) yields it back. Unlike other examples
here we have $T^2 \ne P^2\ne 1$, whereas the results (16) are met. We find
that P and T commute; C and P do not commute. We get $P C \ne (C P)^{-1}
=\eta_+=\eta.$ When $x=1$, the scenario for Hermiticity can be observed.
\section {Conclusions}
\par The theorem stated and proved in section III  adds an important result
in matrix algebra [8] for constructing a metric(s) $\eta_+=(DD^\dagger)^{-1}$ (6)
where $D$ is the diagonalizing matrix for the pseudo-Hermitian matrix which has
real eigenvalues. The proven positive
definiteness (7) of this metric is of utility while constructing the generalized
P, T, C  and an inner product for a matrix-Hamiltonian which possesses a real spectrum.
\par If X is a symmetry
operator for the Hamiltonian H, i.e. $[X,H]=0$ then the proposed definition of
the inner product as $\langle X \Psi |\eta_+\Psi \rangle$ (19) or even
$\langle X \Psi |\eta\Psi \rangle$ is the most general definition proposed so
far [3-9,16,18-20] when Hamiltonians are PT-symmetric or $\eta$-pseudo-Hermitian.
\par We have examined the approach in [16] to be too simple to work in general.
The approach in [18], {\it sans} its inner product, is found to be correct and
more general. However, our modification  of the definition of T makes it
compatible with the proposed indefiniteness of PT-norm and definiteness of CPT-norm [16].
The examples using several matrix Hamiltonians drawn from our recent [15]
studies on pseudo-Hermiticity have illustrated various contentions explicitly.
The works using non-matrix Hamiltonians and yet making similar claims could be
desirable further.
\par Admittedly, the only properties possessed by C, PT, and CPT  are their
involutions (16) various commutations (18) inner product (19), to strike their
correspondance with the actual ${\cal C, PT, and CPT}$ of Hermitian field theory.
Much deeper connections
and agruments would be required to make claims {\it a la} the conventional
$\cal {CPT}$ invariance [17]. One point that requires emphasis is that
in pseudo-Hermiticity, we are able to construct only {\it three} distinct
involutary operators, which we have designated as P, T and C as against
the conventional ${\cal P, T, C}$ [17]. In this regard, our matrix Hamiltonians
could be useful for further refinements in the theory of C, PT, and CPT invariance. Also
these may be taken as toy models of a futuristic pseudo-Hermitian field
theory.
\section*{References }
\begin{enumerate}
\item C.M. Bender and S. Boettcher, Phys. Rev. Lett. 80 (1998) 5243;
\item M. Znojil, Phys. Lett. A {\bf 259} (1999) 220; {\bf 264} (1999) 108.\\
 H.F. Jones, Phys. Lett. A {\bf 262} (1999) 242. \\
 B. Bagchi and R. Roychoudhury, J. Phys. A : Math. Gen. A {\bf 33} (2000) L1.\\
 G. Levai and M. Znojil, J. Phys. A : Math. Gen. {\bf 33} (2000) 7165.\\
 A. Khare and B.P. Mandal, Phys. Lett. A {\bf 272} (2000) 53.\\
 B. Bagchi and C. Quesne, Phys. Lett. A {\bf 273} (2000) 285. \\
 R. Kretschmer and L. Symanowaski, `The interpretation of quantum mechanical
 models with non-Hermitian Hamiltonians and real spectra', arXive
 quant-ph/0105054 \\
 Z. Ahmed, Phys. Lett. A : {\bf 287} (2001) 295; {\bf 286} (2001) 231.\\
 R.S. Kaushal and Parthasarthi, J. Phys. A : Math. Gen. {\bf 35} (2002) 8743.\\
 Z. Ahmed,`Discrete symmetries, Pseudo-Hermiticity and pseudo-unitarity', in
DAE (India) symposium on Nucl. Phys. Invited Talks eds., A.K.Jain and A. Navin,
vol {\bf 45 A} (2002) 172.
\item C. M. Bender, G. V. Dunne, P. N. Meisinger, Phys. Lett. A {\bf 252} (1999) 253.
\item Z. Ahmed, Phys. Lett. A : {\bf 282} (2001) 343.
\item Z. Ahmed, `A generalization for the eigenstates of complex PT-invariant
 potentials with real discrete eigenvalues' (unpublished) (2001).
\item B. Bagchi, C. Quesne and M. Znojil, Mod. Phys. Lett. A{\bf 16} (2001) 2047.
\item G.S. Japaridze, J. Phys. A : Math. Gen. {\bf 35} (2002) 1709.
\item
 R. Nevanlinna, Ann. Ac. Sci. Fenn. {\bf 1} (1952) 108; {\bf 163} (1954) 222.\\
 L.K. Pandit, Nouvo Cimento (supplimento) 11 (1959) 157.\\
 E.C.G. Sudarshan, Phys. Rev. {\bf 123} (1961) 2183.\\
 M.C. Pease III, {\it Methods of matrix algebra} (Academic Press, New York, 1965)\\
 T.D. Lee and G.C. Wick, Nucl. Phys. B {\bf 9} (1969) 209.\\
 F.G. Scholtz, H. B. Geyer and F.J.H. Hahne, Ann. Phys. {\bf 213} (1992) 74.
\item P.A.M. Dirac, Proc. Roy. Soc. London {A 180} (1942) 1.\\
 W. Pauli, Rev. Mod. Phys. {\bf 15} (1943) 175.\\
 T.D. Lee, Phys. Rev. {\bf 95} (1954)  1329.\\
 S.N. Gupta, Phys. Rev. {\bf 77} (1950) 294.\\ K. Bleuler, Helv. Phys. Act.
 {\bf 23} (1950) 567.
\item A. Mostafazadeh, J. Math. Phys. {\bf 43} (2002) 205;{\bf 43} (2002) 2814;
{\bf 43} (2002) 3944.
\item Z. Ahmed, Phys. Lett. A {\bf 290} (2001) 19; {\bf 294} (2002) 287.
\item L. Solombrino,`Weak pseudo-Hermiticity and antilinear commutant', arXive
math-ph/0203101.
\item G,Solarici, J. Phys. A: Math. Gen. {\bf 35} (2002) 7493
\item T,V. Fityo, J. Phys. A : Math. Gen. {\bf 35} (2002) 5893.
\item Z. Ahmed and S.R. Jain, ``Pseudo-unitary symmetry and the Gaussian pseudo-unitary
ensemble of random matrices", arXiv quant-ph/0209165 (also submitted to Phys.
Rev. E.)\\
Z. Ahmed and S.R. Jain, ``Gaussian ensembles of $2 \times 2$ pseudo-Hermitian
random matrices" to appear in J. Phys. A: Math. Gen.  (2003, The special issue
on Random Matrices).\\
Z. Ahmed, ``An ensemble of non-Hermitian Gaussian-random $2\times 2$ matrices
admitting the Wigner surmise'' to appear in Phys. Lett. A (PLA 12155) 2003.
\item C.M. Bender, D.C.Brody and H.F.Jones, ``Complex extension of quantum
mechanics'', arXive quant-ph/2010076, Phys. Rev. Lett. {\bf 89} (2002) 270401.
\item R.F. Streater and A.S. Wightman, ${\cal PCT}$, {\em Spin and Statistics, and
all that} (Benjamin, New York, 1964).
\item A. Mostafazadeh, `Pseudo-Hermiticity and Generalized PT-and CPT-symmetries'
arXive math-ph/0209018 (to appear in J. Math, Phys.).
\item C.M. Bender, P.N.Meisinger and Q. Wang, ``All Hermitian Hamiltonians
have parity'', arXive quant-ph/0211123, J. Phys. A : Math. Gen. {\bf 36} (2003) 1029.
\item Z. Ahmed, `P, T, PT, and CPT invariance of Hermitian Hamiltonians', arXive
quant-ph/0302084.
\end{enumerate}
\end{document}